\documentclass[preprint,12pt]{elsarticle}

\usepackage{graphicx} %
\usepackage{multirow}%
\usepackage{amsmath,amssymb,amsfonts}%
\usepackage{amsthm}%
\usepackage{mathrsfs}%
\usepackage[title]{appendix}%
\usepackage{xcolor}%
\usepackage{textcomp}%
\usepackage{manyfoot}%
\usepackage{booktabs}%
\usepackage{algorithm}%
\usepackage{algorithmicx}%
\usepackage{algpseudocode}%
\usepackage{listings}%
\usepackage{array}

\journal{Physica A}

\begin{document}

\begin{frontmatter}

\title{Why We Experience Society Differently: Intrinsic Dispositions as Drivers of Ideological Complexity in Adaptive Social Networks}

\author[a]{Akshay Gangadhar} 
\ead{agangadhar@binghamton.edu}
\author[a,b]{Hiroki Sayama} 
\ead{sayama@binghamton.edu}

\affiliation[a]{organization={Binghamton Center of Complex Systems, Binghamton University, State University of New York},
            addressline={}, 
            city={Binghamton},
            postcode={13902-6000}, 
            state={NY},
            country={USA}}

\affiliation[b]{organization={Waseda Innovation Lab},
            addressline={Waseda University}, 
            city={Tokyo},
            postcode={169-8050}, 
            country={Japan}}

\begin{abstract}

Understanding the emergence of inequality in complex systems requires attention to both structural dynamics and intrinsic heterogeneity. In the context of opinion dynamics, traditional models relied on static snapshots or assumed homogeneous agent behavior, overlooking how diverse cognitive dispositions shape belief evolution. While some recent models introduce behavioral heterogeneity, they typically focus on macro-level patterns, neglecting the unequal and individualized dynamics that unfold at the agent level.
In this study, we analyze an adaptive social network model where each agent exhibits one of three behavioral tendencies—homophily, neophily (attention to novelty), or social conformity—and measure the complexity of individual opinion trajectories using normalized Lempel-Ziv (nLZ) complexity. We find that the resulting dynamics are often counterintuitive—homophilic agents, despite seeking similarity, become increasingly unpredictable; neophilic agents, despite pursuing novelty, exhibit constrained exploration; and conformic agents display a two-phase trajectory, transitioning from early suppression of variability to later unpredictability. More fundamentally, these patterns remain similar across diverse network settings, suggesting that internal behavioral dispositions — rather than external environment alone — play a central role in shaping long-term opinion unpredictability. The broader implication is that individuals’ experiences of ideological volatility, uncertainty, or stability are not merely environmental, but may be endogenously self-structured through their own cognitive tendencies. 
These results establish a novel individual-level lens on opinion dynamics, where the behavioral identity of agents serves as a dynamical fingerprint in the evolution of belief systems, and gives rise to persistent disparities in dynamical experience within self-organizing social systems, even in structurally similar environments.

\end{abstract}


\end{frontmatter}

\section{Introduction}
In today’s highly interconnected world, the increasing prevalence of extreme opinions and ideological polarization has become a defining characteristic of modern social systems \cite{conover2011political,prior2013media,morales2015measuring,manrique2018generalized,cole2011vegaphobia,reilly2016gluten,johnson2019health}. While such polarization is commonly attributed to external triggers—such as political events or media manipulation—it can also emerge as a spontaneous outcome of social dynamics driven by human-centric factors \cite{sayama2020extreme,sayama2022social}. The advent of modern information communication technology has further intensified these dynamics by enabling preferential selection of information sources, often leading to the formation of “social bubbles” and echo chambers \cite{sayama2020enhanced,nikolov2015measuring,spohr2017fake}. As societies grapple with issues such as globalization, immigration, and cultural diversity, individuals and groups may retreat into more homogeneous communities as a means of preserving their sense of identity and security. 

To understand the formation of extreme opinions and the structural evolution of fragmented societies, researchers have studied phase transitions between connected and fragmented states in adaptive social networks \cite{holme2006nonequilibrium,zanette2006opinion,bohme2011analytical,gross2009adaptive}. Within these adaptive social networks, individuals adjust their social ties and opinions based on local interactions, producing complex, self-organizing patterns of social fragmentation and ideological entrenchment. Traditional bounded confidence models have demonstrated that polarization can emerge even under moderate openness to differing views \cite{rainer2002opinion,liang2013opinion}. Prior research has shown that heterogeneity in update policies can significantly influence the formation of extremist communities and network connectivity \cite{bennett2022exploiting,bullock2023agent}.

A key limitation of much of the existing literature on opinion dynamics lies in its reliance on static snapshots of network states or opinion distributions. These approaches—often based on modularity, clustering, or structural detection of polarization—offer only a momentary view of ideological divisions, failing to account for how opinions and network structures evolve over time. In response, dynamic models of social networks were developed to capture temporal evolution; however, these models typically assumed homogeneous behavioral rules across agents, neglecting the diversity of cognitive dispositions that shape real-world social behavior \cite{sayama2020enhanced,bernardo2022finite}.

To address this, more recent studies introduced behavioral heterogeneity into dynamic opinion models \cite{sayama2022social,chen2019modeling,li2023bounded}. While this marked an important advancement, such analyses have largely focused on aggregate or system-level outcomes, such as overall polarization or network connectivity. These macro-scale metrics, while informative, obscure the potentially rich and unequal dynamical behaviors at the level of individual agents.

This study addresses that gap by explicitly analyzing individual-level opinion dynamics within a behaviorally heterogeneous adaptive social network. By tracking the temporal unpredictability of each node’s opinion trajectory, we uncover how internal behavioral dispositions—not just external network structure—govern the complexity of ideological evolution. This focus on individual dynamical fingerprints represents a novel approach to understanding polarization and social fragmentation in adaptive systems.

In this study, we focus on normalized Lempel–Ziv (nLZ) complexity, which quantifies the compressibility of entire opinion trajectories and is inherently sensitive to long-range temporal structure \cite{lempel1976complexity,zhang2009normalized}. This choice is motivated by our interest in characterizing differences in opinion dynamics that develop cumulatively over time, rather than by instantaneous or short-term fluctuations. In contrast, measures based on short temporal windows or local statistics primarily capture local ordering and variability, and may therefore be less informative when distinguishing features of the dynamics that—if present—emerge only over longer time scales.  

We study the adaptive social network model developed in  \cite{sayama2020extreme} that explains how extreme ideas can spontaneously arise based on the interplay between three factors: homophily, attention to novelty (neophily), and social conformity. Here, it was found that, while homophily reinforces segregation by strengthening ties between like-minded individuals, neophily encourages diversity by fostering interactions with those who hold novel viewpoints. Social conformity further complicates these dynamics, as it can either promote ideological convergence or drive extremization depending on network conditions. nLZ complexity, here, can be  understood as the characterization of the unpredictability in the opinion-evolution trajectory of nodes in this adaptive social network. A node undergoing a more rapid, diverse, and extreme set of opinions can be said to have an experience of a more dynamic and unpredictable nature as compared to that of a node undergoing a more constrained and redundant set of opinions in the network. The use of this measure, therefore, allows us to understand the relationship between the inherent unpredictability in opinion trajectories of nodes in these adaptive networks with respect to their internal dispositions (homophily, neophily, and social conformity).  

By analyzing nLZ trends across three experimental scenarios (described in the following section), our objective is to identify faction-specific complexity patterns of opinion trajectories and examine how node-level behavioral tendencies shape the dynamical complexity of opinion evolution.

\section{Methods}
\subsection{The model} 
The original model \cite{sayama2020extreme} describes an adaptive social network with \textit{n} nodes, with each node \textit{i} having its own opinion $x_i \in \mathbb{R}$. The edge weight $w_{ij}$ denotes flow of information from source node \textit{j} to target node \textit{i}. The dynamics of the network, as indicated by evolving nodal opinions and edge weights are described by the following set of equations:

\begin{align}
    \dfrac{dx_i}{dt} & =  c(\langle x \rangle_i-x_i) + \epsilon \\
    \dfrac{dw_{ij}}{dt} & =  h F_h (x_i,x_j) + a F_a (\langle x \rangle_i,x_j) \\ 
    \langle x \rangle_i & =  \dfrac{\sum_j w_{ij} x_j}{\sum_j w_{ij}}
\end{align}

Here, $\langle x \rangle_i$ is the weighted average of opinions in node \textit{i}'s neighborhood, i.e., the social norm as perceived by node \textit{i}. \textit{c} and $\epsilon$ are social conformity and noise terms that impact the evolution of node \textit{i}'s opinion. In the second equation, \textit{h} and \textit{a} are the homophily and attention to novelty factors that impact the change in the edge weight for the edge directed from node \textit{j} to node \textit{i}, in other words, the degree of closeness between the two nodes \textit{i} and \textit{j}. Here, $F_h$ and $F_a$ are behavioral functions that determine the degree of change in edge weights. These functions are adopted from the original study and are described below: 

\begin{align}
    F_h(x_i, x_j) & = \theta_h - |x_i - x_j| \\ 
    F_a(\langle x \rangle_i,x_j) & = |\langle x \rangle_i - x_j| - \theta_a
\end{align}

$\theta_h$ and $\theta_a$ are constants that define the function values when the arguments are equal. Essentially, $F_h$ designates the strengthening of edge connecting \textit{j} to \textit{i} when nodes \textit{i} and \textit{j} have similar opinions. $F_a$ designates the strengthening of the same edge when node \textit{j} has an opinion different from that of the perceived social norm of node \textit{i} (i.e., novelty). In these simulations, $w_{ij}$ is bound to be \textit{non-negative} and the edge weights are set to zero when they go negative. 

\begin{figure}
    \centering
    \includegraphics[width=\linewidth]{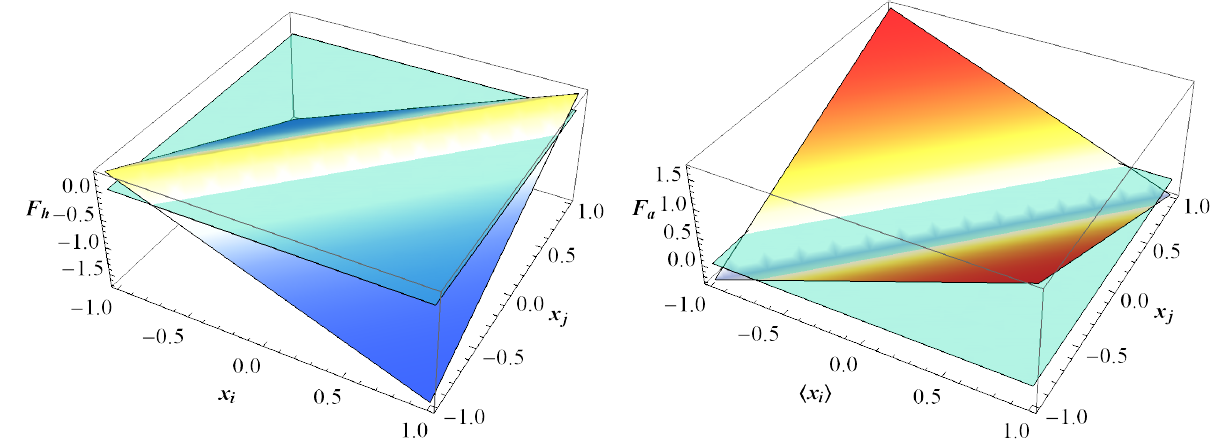}
    \caption{Functional shapes for the behavioral functions $F_h$ (left) and $F_a$ (right) for $\theta_h = \theta_a = 0.3$ (adopted from \cite{sayama2020extreme}). The cyan planes indicate zero level.} 
    \label{fig:enter-label}
\end{figure}

\subsection{Lempel-Ziv Complexity}
Lempel-Ziv (LZ) complexity measures a sequence's predictability by assessing its compressibility \cite{lempel1976complexity}. It identifies repeated substrings within the sequence; the more unique substrings found, the greater the unpredictability. High LZ sequences are harder to compress as they contain fewer repeating patterns, while low LZ sequences compress more readily due to greater redundancy. Therefore, the efficiency of compression reflects the sequence's richness: more compressible sequences signify greater \textit{predictability}, while resistance to compression indicates greater \textit{unpredictability}. 

To calculate the LZ complexity, we traverse a string from left to right, adding each new substring to a dictionary. If a substring repeats, we extend the search span to find a new unique substring. For instance, in ``01100101101100100110", we encounter ``0" and ``1," each added to the dictionary. Then, upon encountering another ``1" we extend the search span to two, to ``10", which is added. Continuing in this manner, the string is parsed as 0\textbullet{}1\textbullet{}10\textbullet{}010\textbullet{}1101\textbullet{}100100\textbullet{}110~\cite{shmulevich2005eukaryotic}. Note that the final substring may not be unique. The LZ complexity of this sequence is equal to the dictionary's length, representing the number of unique substrings. In this case, it is 7. 

In this study, we use the normalized Lempel-Ziv (nLZ) complexity which, essentially, makes the measure independent of the length of the sequence (since LZ is clearly correlated with sequence length) \cite{zhang2009normalized}. This is done as:

\begin{equation}
    \text{\textit{nLZ}} = \dfrac{LZ}{\left(\dfrac{\text{\textit{n}}}{\text{log \textit{n}}}\right)}
\end{equation}

where \textit{n} is the length of the sequence.  

This is justified, since it has been shown that the LZ complexity for a sequence of uniformly distributed characters approaches $\dfrac{\text{\textit{n}}}{\text{log \textit{n}}}$ as \textit{n} becomes arbitrarily large \cite{lempel1976complexity}.

For our model, we compute the nLZ complexity for each node's time series of opinions (until a given time step) and average this across all nodes of a given type (depending on the update policy - highly homophilic/neophilic/conformic) to obtain a measure of the predictability of opinion trajectory for nodes of that faction. 

\subsection{Discretization of the model}
Two forms of discretization are employed in our model: temporal discretization and opinion discretization.

\textit{Temporal discretization.}  
The underlying ordinary differential equation (ODE) model is discretized in time using a forward Euler scheme. Node states are updated at discrete time steps of size $\Delta t = 0.1$. Let $t$ denote the discrete time index. Then $t=0$ corresponds to simulation time $0$, $t=10$ corresponds to simulation time $1$, $t=50$ corresponds to simulation time $5$, and so on.

\textit{Opinion discretization.}  
The normalized Lempel--Ziv (nLZ) complexity is defined for discrete symbolic sequences, whereas the opinion states in our model evolve continuously. To enable the computation of nLZ, continuous opinion values are discretized using a uniform binning scheme with bin width $\Delta$, centered around zero. Each opinion value is mapped to the midpoint of the bin in which it falls, yielding a discrete symbolic time series of opinions for each node.

Because discretization can, in principle, introduce artifacts that affect measured complexity, we explicitly examine the sensitivity of our results to the choice of bin width. To this end, we conduct a bin-sensitivity analysis in which nLZ trajectories are computed across a range of bin sizes, $\Delta \in [0.5,\,1.0]$ (Figures S1-S5). Within this range, we observe only minor quantitative variations in the nLZ curves. Crucially, the qualitative dynamical signatures that form the core contributions of this study—monotonic growth under homophily, biphasic dynamics under conformity, and stable, bounded complexity under neophily—remain invariant across all tested bin sizes. For bin widths substantially smaller than this range, discretization produces overly fine symbolic sequences that inflate nLZ values. Conversely, for bin widths substantially larger than this range, discretization becomes overly coarse, leading to diminished nLZ values and a loss of meaningful temporal structure. Accordingly, all results reported in this paper use a representative bin width of $\Delta = 0.75$.

\subsection{Experimental setup}
Three scenarios were devised, each an extension of the previous, to explore the space of possibilities in the adaptive social network model. These are described below.

\begin{itemize}
    \item \textit{Scenario 1}: Create ``pure" networks comprising of agents, all biased toward one of the update policies.   
    \begin{itemize}
        \item Homophilic networks: \textit{a}, \textit{c} $\sim$ $N(0.05, 0.025)$; \textit{h} $\sim$ $N(0.25,0.025)$
        \item Neophilic networks: \textit{h}, \textit{c} $\sim$ $N(0.05, 0.025)$; \textit{a} $\sim$ $N(0.25,0.025)$
        \item Conformic networks: \textit{h}, \textit{a} $\sim$ $N(0.05, 0.025)$; \textit{c} $\sim$ $N(0.25,0.025)$  
    \end{itemize}
    \item \textit{Scenario 2}: Generate 50-50 ``dual faction" networks comprising of two equally proportioned factions each biased toward one of the three update policies.   
    \begin{itemize}
        \item This creates three types of networks where we observe network evolution as a result of interactions between two competing factions in the population (\textit{h vs a, a vs c, c vs h}).
        \item Node parameterization is based on the same distributions as Scenario 1 (depending on the faction).
    \end{itemize}
    \item \textit{Scenario 3}: Create networks of mixed agents (with uniformly distributed node parameters).
    \begin{itemize}
        \item \textit{a, c, h} $\sim$ $U(0.05, 0.3)$
    \end{itemize}
\end{itemize}

In Scenario 1, pure networks are created where all agents share the same bias—resulting in homophilic, neophilic, or conformic networks. In Scenario 2, dual faction networks are formed with two competing biases (\textit{h vs a, a vs c, c vs h}) to study how interactions between two factions affect opinion complexity trends. Finally, Scenario 3 uses a mixed setup with uniformly distributed biases, providing a fully heterogeneous environment. This approach systematically examines how opinion complexity trends vary from homogeneous settings (where all agents are of the same type) to heterogeneous settings (comprising of a mixed pool of different types of agents). 

All parameter distributions have been adopted from the original study (exploring the impact of heterogeneity in update policies) to ensure consistency~\cite{bullock2023agent}. Each network consists of 300 nodes. 10 networks each were generated for the three scenarios (10 each for the three pure networks of Scenario 1; 10 each for the three dual faction networks of Scenario 2; 10 mixed networks for Scenario 3), and simulated for 3000 time-steps.

\section{Results} 
\begin{figure}[h]
    \centering
    \includegraphics[width=1\linewidth]{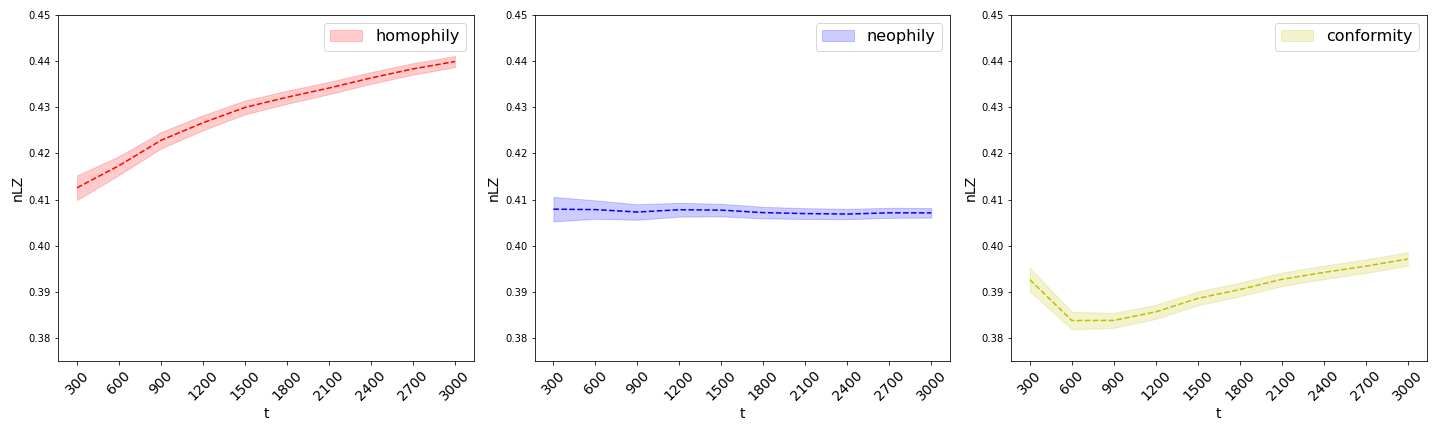}
    \caption{\textit{nLZ vs t} for the three pure networks in Scenario 1. Homophilic (left) vs Neophilic (center) vs Conformic (right) node nLZ complexities (with 95\% confidence bounds) are shown in distinct colors. The nLZ value is computed over increasing time intervals to capture changes in opinion trajectorial unpredictability over time.}
    \label{fig:scenario1}
\end{figure}

\begin{figure}
    \centering
    \includegraphics[width=1\linewidth]{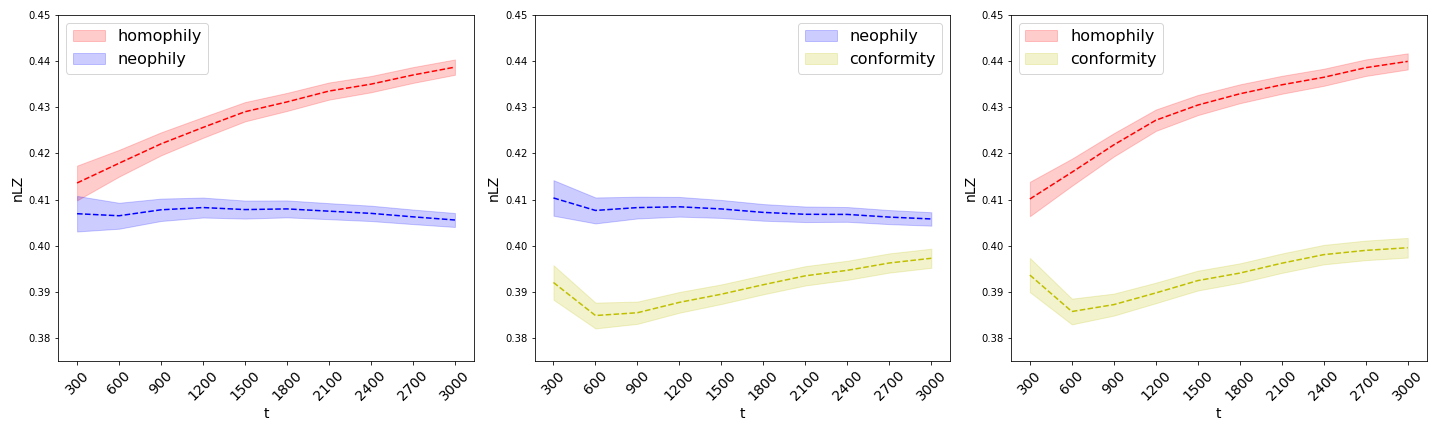}
    \caption{\textit{nLZ vs t} for the three 50-50 networks in Scenario 2 (with 95\% confidence bounds). Homophily vs Neophily (left); Neophily vs Conformity (center); Conformity vs Homophily (right).}
    \label{fig:scenario2}
\end{figure}

\begin{figure}
    \centering
    \includegraphics[width=0.5\linewidth]{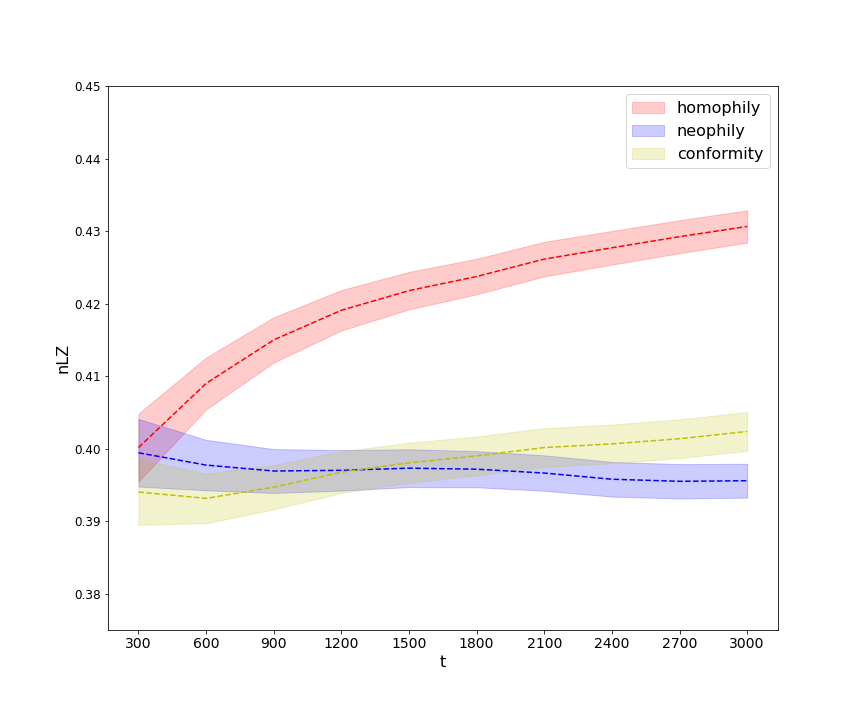}
    \caption{\textit{nLZ vs t} for the mixed networks in Scenario 3 (with 95\% confidence bounds).}
    \label{fig:scenario3}
\end{figure}

Figures~\ref{fig:scenario1},~\ref{fig:scenario2}, and~\ref{fig:scenario3} illustrate the evolution of nLZ complexity over time across all three experimental scenarios. nLZ complexity is computed for each node's opinion trajectory over incrementally increasing time windows (from $t=0$ to $t=3000$ in steps of 300), and results are averaged within each node type to assess characteristic patterns.

For each independent simulation replicate, we first compute the mean nLZ trajectory by averaging node-level nLZ values within a faction at each time point. Confidence bands are then obtained by bootstrapping these replicate-level mean trajectories across independent replicates, yielding 95\% confidence intervals for the mean nLZ as a function of time.

\subsection{Distinctive complexity profiles by node type}
The analysis reveals behavior-specific complexity signatures for homophilic, neophilic, and conformic nodes across all scenarios.  

Homophilic nodes consistently exhibit a steadily increasing nLZ trend, indicating progressively more unpredictable opinion dynamics. While this appears counterintuitive—given that homophilic individuals are presumed to seek stability through similarity—the model reveals that their persistent reinforcement of like-mindedness may paradoxically lead to long-term instability. Local fluctuations, minor disagreements, and evolving neighborhood structures introduce compounding variability, which could drive up the unpredictability of their opinion shifts.

In contrast, neophilic nodes maintain a low and stable nLZ complexity profile over time across all network settings. Although these nodes actively seek novelty, their opinion trajectories stabilize early in the simulation, resulting in bounded and persistent levels of unpredictability. This behavior suggests that novelty-seeking may lead to dynamically constrained exploration, where adaptive interactions limit the continual generation of new symbolic patterns. Consequently, neophilic agents exhibit structured yet non-convergent dynamics, maintaining consistent nLZ values despite ongoing exposure to diverse opinions and a continuously evolving social environment.

Conformic nodes exhibit a two-phase nLZ pattern. Early in the simulation, nLZ tends to decrease or remain stable, consistent with short-term suppression of variability under conformity. At later times, nLZ increases, indicating growing unpredictability as the adaptive network evolves. While the underlying mechanisms remain unclear, this behavior may reflect an initial promotion of local homogeneity followed by the gradual accumulation of inconsistencies that erode predictability over longer timescales. This dual-phase behavior suggests that social conformity can foster both short-term homogeneity and long-term instability, depending on the stage of network evolution.

\subsection{Counterintuitive behavioral dynamics}
The behavioral patterns observed in the model are frequently unintuitive and challenge common assumptions about the psychological correlates of homophily, novelty-seeking, and conformity. Importantly, the interpretations offered below are phenomenological and intended to suggest plausible explanations rather than definitive mechanisms:
\begin{itemize}
    \item Homophilic nodes, despite a drive for consistency and alignment, exhibit increasingly unpredictable opinion trajectories over time. This counterintuitive trend suggests that persistent self-reinforcement within homogeneous neighborhoods may, under adaptive dynamics, contribute to growing variability rather than long-term stability.
    \item Neophilic nodes, which might be expected to display highly erratic dynamics due to continual novelty-seeking, instead maintain stable and bounded levels of complexity. One possible interpretation is that sustained exposure to diverse opinions leads to regulated or constrained exploration, limiting the accumulation of new symbolic patterns.
    \item Conformic nodes display a two-phase complexity pattern, with early-time predictability followed by increasing unpredictability at later stages. This behavior may reflect short-term suppression of variability through conformity, which gradually gives way as adaptive interactions and local inconsistencies accumulate over time.
\end{itemize}

\subsection{Internal behavioral rules as drivers of complexity}

A striking observation in this study is the similarity of nLZ complexity trends across all three experimental scenarios—pure, dual-faction, and fully heterogeneous networks. Across these settings, nodes grouped by behavioral predisposition (homophily, neophily, or conformity) tend to exhibit similar qualitative complexity patterns, even as the surrounding social environment varies substantially. Within the context of this adaptive network model, this suggests a strong association between internal behavioral rules and the long-term complexity of individual opinion trajectories.

Importantly, these results do not, by themselves, establish a causal relationship. The observed alignment between node type and complexity profile may reflect correlations induced by the model’s structure, parameterization, or coevolutionary dynamics. Nevertheless, the persistence of behavior-specific nLZ signatures across diverse network compositions motivates the hypothesis that internal behavioral predispositions play a central role in shaping individual-level experiences of opinion unpredictability, potentially outweighing the influence of external social context within this modeling framework.

Viewed through this lens, the model raises an intriguing possibility: that the perceived complexity or volatility of an individual’s opinion dynamics may be systematically linked to their own behavioral tendencies, rather than being solely a product of the social environment they inhabit. For example, within the model, homophilic nodes tend to experience increasing unpredictability over time even when embedded in different network compositions, suggesting that such volatility may emerge endogenously from their interaction rules rather than exogenously from environmental instability.

While any extrapolation beyond the model must be made cautiously, this perspective resonates with broader discussions in psychology and sociology regarding the interplay between individual dispositions and social experience. The adaptive social network studied here provides a simplified and stylized setting in which such ideas can be explored, not as definitive explanations, but as testable hypotheses. In the following subsection, we therefore move beyond observational consistency and perform explicit causal intervention tests to assess whether internal behavioral rules exert a direct influence on opinion complexity dynamics.

\subsection{Causal intervention test: isolating internal disposition from external context}

The results presented in the previous sections demonstrate that symbolic complexity signatures, as measured by nLZ complexity, exhibit similar qualitative patterns across three increasingly realistic scenarios: homogeneous populations (Scenario~1), dual faction regimes (Scenario~2), and fully heterogeneous environments (Scenario~3). In particular, these analyses suggest that internal behavioral dispositions---homophily, neophily, and conformity---play a dominant role in shaping the unpredictability of individual opinion trajectories.

However, the analyses so far remain fundamentally correlational. Although the persistence of nLZ signatures across scenarios provides strong evidence of robustness, it does not by itself establish a \emph{causal} relationship between internal dispositions and observed complexity patterns. In principle, the observed signatures could still arise indirectly from environmental structure, network effects, or collective feedback mechanisms rather than from node-level behavioral parameters themselves.

To directly address this limitation, and to test whether internal dispositions can override external circumstances in a causal sense, we introduce an explicit intervention-based experiment.

\subsubsection{Intervention rationale}

Our core causal question is the following: \emph{if the internal behavioral disposition of a single node is altered while the surrounding social environment is held fixed, does the symbolic complexity of that node's opinion trajectory change in a predictable and systematic manner?}

This formulation mirrors classical intervention logic: the system is allowed to evolve naturally until a specified time, at which point a controlled manipulation is applied to an internal parameter, while all external influences---including network topology, edge weights, and neighboring node states---remain unchanged. Any subsequent divergence in complexity can therefore be attributed directly to the internal intervention.

This approach is particularly important in heterogeneous environments (Scenario~3), where the external context already contains substantial variability. Demonstrating causal effects under these conditions provides a stringent test of whether internal dispositions genuinely drive complexity rather than merely correlating with favorable environments.

\subsubsection{Experimental design}

For each scenario, we simulate opinion dynamics on a directed, weighted network for a total duration of $T=3000$ time steps. At predetermined intervention times
\[
t^* \in \{600, 900, 1200\},
\]
we store a complete snapshot of the network state, including all node opinions and edge weights.

From each snapshot, we generate two trajectories for a selected focal node:
\begin{enumerate}
    \item a \emph{baseline} continuation, in which the node's internal parameters remain unchanged;
    \item an \emph{intervention} continuation, in which the node's internal behavioral parameters are modified to represent a different disposition (e.g., homophilic $\rightarrow$ conformic).
\end{enumerate}

Crucially, both continuations evolve forward from the \emph{same} network state at $t^*$, ensuring that the surrounding social environment is identical in both cases. The intervention therefore isolates the causal effect of internal disposition alone. 

\paragraph{Intervention parameter assignment}
Across both Scenario~1 (homogeneous) and Scenario~3 (heterogeneous), interventions are implemented by reassigning the focal node’s internal parameters using the same sampling procedure employed in Scenario~1. Specifically, when a node is switched to a target disposition, its parameters are resampled from normal distributions that enforce strong dominance of the target behavior. For example, a switch to a homophilic disposition assigns
\[
h \sim \mathcal{N}(0.25, 0.025), \quad
a \sim \mathcal{N}(0.05, 0.025), \quad
c \sim \mathcal{N}(0.05, 0.025),
\]
with analogous assignments used for neophilic and conformic targets. This ensures that each intervention represents an extreme and unambiguous change in internal behavioral regime, independent of the node’s prior parameter values.

\paragraph{Selection of focal nodes}
In Scenario~1, focal nodes are selected directly from homogeneous populations in which all nodes share the same dominant disposition. In Scenario~3, where node parameters are drawn heterogeneously, focal nodes are selected using an explicit dominance criterion: a node is classified as homophilic, neophilic, or conformic if its corresponding parameter is the largest among $\{h,a,c\}$ and exceeds a threshold $\tau$ (set to $\tau=0.25$ in our analysis). This criterion ensures that interventions are applied only to nodes exhibiting a sufficiently strong baseline disposition, thereby preserving interpretability of the transition being tested.

nLZ complexity is evaluated at increasing time intervals along each node's opinion trajectory. Specifically, nLZ is computed on discretized trajectories of length
\[
[0,300], [0,600], \ldots, [0,3000],
\]
This allows us to track how intervention-induced changes in symbolic complexity accumulate over time.

\subsubsection{Quantifying the intervention effect}

To move beyond visual comparison of complexity trajectories, we define a scalar post-intervention effect size for each independent simulation run. For a fixed intervention time $t^*$, we compute
\[
\Delta
=
\left\langle
\mathrm{nLZ}_{\text{intervention}}(t)
-
\mathrm{nLZ}_{\text{baseline}}(t)
\right\rangle_{t \ge t^* + \Delta t},
\]
where the average is taken over all time steps strictly after the intervention (with $\Delta t=300$ ensuring separation from transient effects).

This quantity measures the mean shift in nLZ complexity induced by the intervention, aggregated over the post-intervention period. Positive values of $\Delta$ indicate an increase in unpredictability due to the internal change, while negative values indicate a suppression of complexity.

For each transition and intervention time, we analyze the distribution of $\Delta$ across independent simulation runs using:
\begin{itemize}
    \item the mean and 95\% confidence interval of $\Delta$,
    \item a one-sample $t$-test against zero,
    \item Cohen's $d$ as a standardized effect size.
\end{itemize}

While statistical significance establishes the reliability of the effect, Cohen's $d$ provides a scale-free measure of its magnitude relative to natural variability, enabling meaningful comparison across scenarios and transitions.

\subsubsection{Causal intervention test results}

\begin{figure}[h!]
    \centering
    \includegraphics[width=0.49\linewidth]{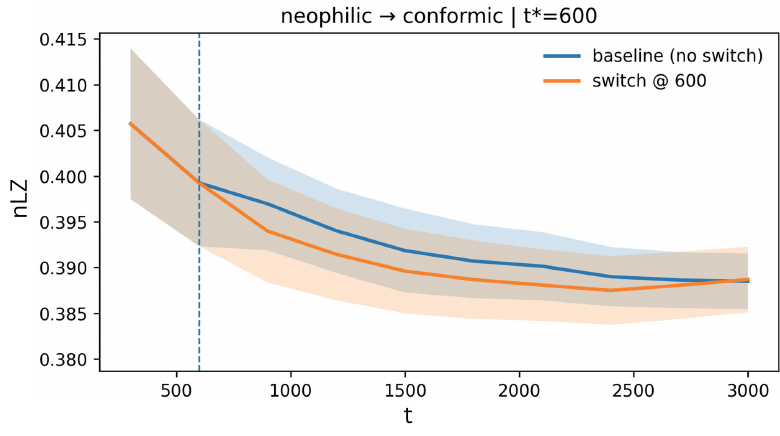}
    \includegraphics[width=0.49\linewidth]{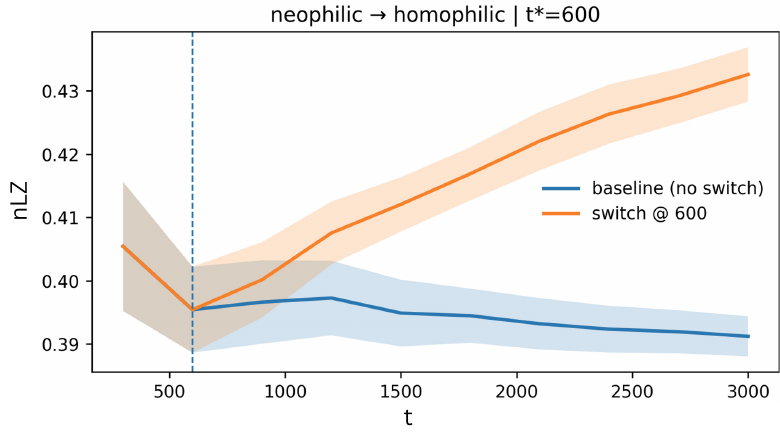}\\
    \includegraphics[width=0.49\linewidth]{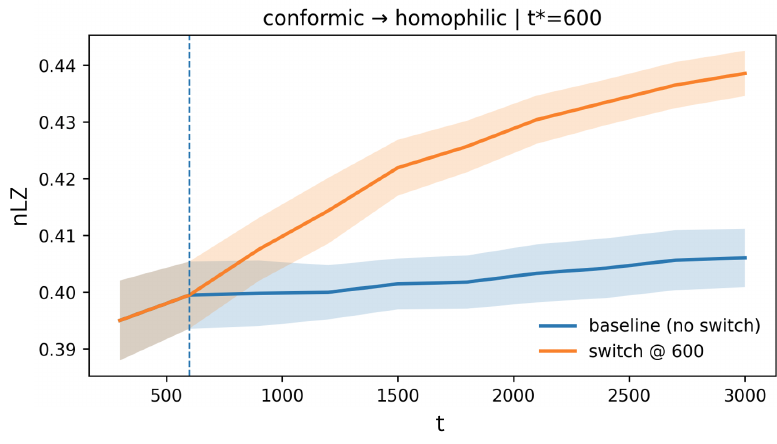}
    \includegraphics[width=0.49\linewidth]{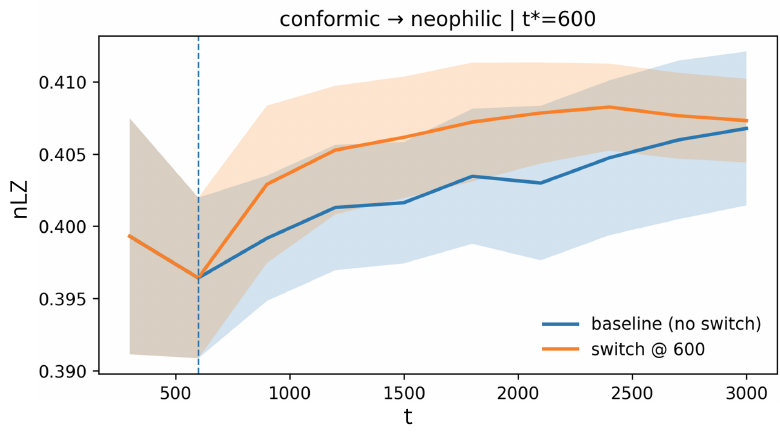}\\
    \includegraphics[width=0.49\linewidth]{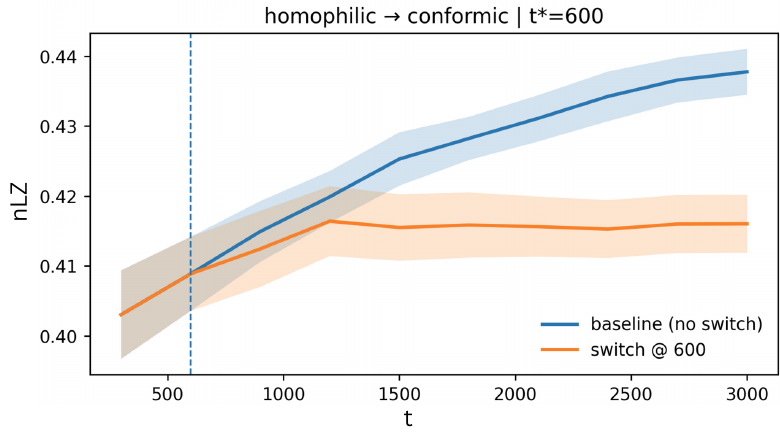}
    \includegraphics[width=0.49\linewidth]{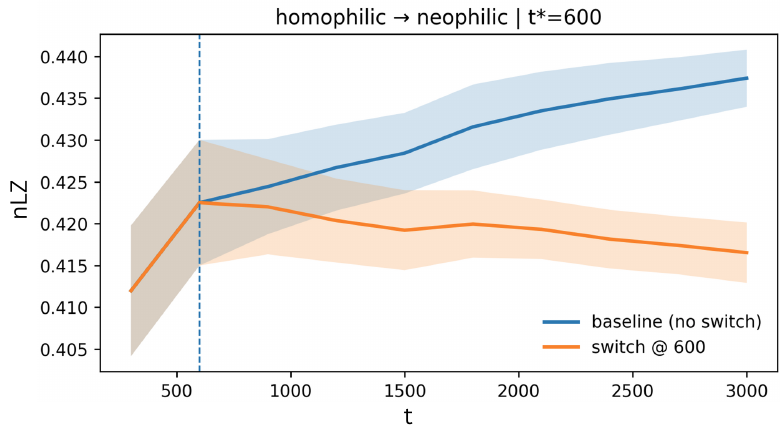}
    \caption{
    Causal intervention results for Scenario~3 (heterogeneous environment) at intervention time \(t^*=600\).
    For each transition, the nLZ complexity of the focal node's opinion trajectory is shown for the baseline (no intervention; in blue) and intervention (in orange) cases, averaged across nodes and runs with 95\% confidence intervals.
    }
    \label{fig:causal_s3}
\end{figure}

\paragraph{Causal effects in heterogeneous environments}
Figure~\ref{fig:causal_s3} shows the causal intervention results for Scenario~3 (heterogeneous environments) at \(t^*=600\) for all six behavioral transitions. Despite substantial environmental variability, switching the internal disposition of a single node produces systematic and directionally consistent deviations in nLZ relative to the no-intervention baseline in most cases.

Table~\ref{tab:causal_s3_summary} provides a quantitative summary of these effects. Several transitions exhibit statistically significant post-intervention shifts with moderate to large effect sizes, demonstrating that internal behavioral dispositions exert a measurable causal influence even when external conditions are heterogeneous.  

Complete results for all intervention times in Scenario~3, as well as the corresponding causal intervention results for Scenario~1, are provided in the Supplementary Material (Tables~S1 and~S2; Figures~S6 and~S7). These additional analyses are consistent with the trends observed here, exhibiting the same directional structure of post-intervention nLZ shifts with effect magnitudes that vary systematically with intervention time and environmental heterogeneity.

\paragraph{Interpretation}
Taken together, these results provide direct causal evidence that internal behavioral dispositions are not merely correlated with symbolic complexity signatures but actively shape the unpredictability of opinion trajectories. Even in heterogeneous environments, altering a single node’s internal parameters produces systematic post-intervention shifts in nLZ, demonstrating a causal influence that persists despite substantial environmental variability.

At the same time, the intervention analysis reveals that transitions between neophily and conformity do not produce statistically significant changes in long-term unpredictability. In other words, within heterogeneous environments, switching a node between these two dispositions does not lead to a discernible divergence in its nLZ trajectory relative to the no-intervention baseline. This observation is consistent with prior studies showing that, despite their distinct microscopic mechanisms, both neophilic and conformic tendencies can drive networks toward similar macroscopic outcomes, such as homogenized opinions and well-connected interaction structures \cite{sayama2020extreme,sayama2022social}. From this perspective, the lack of separation observed here may reflect a genuine dynamical similarity at the macro level rather than an absence of causal influence.

Overall, these findings reinforce the central claim of this work: internal behavioral dispositions exert a causal influence on node-level unpredictability, even in heterogeneous social environments, while also highlighting that different microscopic tendencies may converge to comparable macroscopic signatures under certain conditions.

\begin{table}[]
\centering
\caption{
Causal intervention summary for Scenario~3 (heterogeneous environment) at \(t^*=600\).
Reported values are the post-intervention mean nLZ shift \(\Delta \pm\) 95\% confidence interval, with significance indicated by stars, and Cohen’s \(d\).
Stars denote significance from a one-sample \(t\)-test against zero:
\(*: p<10^{-2}\), \(**: p<10^{-3}\), \(***: p<10^{-4}\), and so on.
}
\label{tab:causal_s3_summary}
\begin{tabular}{lcc}
\\
\toprule
Transition & \(\Delta \pm CI\) & Cohen’s \(d\) \\ 
\midrule 
A $\rightarrow$ C & $-0.00172 \pm 0.00411^{}$ & $-0.16$ \\
A $\rightarrow$ H & $+0.02435 \pm 0.00332^{***********}$ & $+2.87$ \\
C $\rightarrow$ H & $+0.02327 \pm 0.00568^{******}$ & $+1.60$ \\
C $\rightarrow$ A & $+0.00332 \pm 0.00438^{}$ & $+0.29$ \\
H $\rightarrow$ C & $-0.01311 \pm 0.00397^{****}$ & $-1.29$ \\
H $\rightarrow$ A & $-0.01251 \pm 0.00347^{*****}$ & $-1.41$ \\
\bottomrule
\end{tabular}
\end{table}

\section{Conclusions}

In this study, we examined opinion evolution in an adaptive social network model through the lens of normalized Lempel--Ziv (nLZ) complexity, using it as a measure of the temporal unpredictability of individual opinion trajectories. Across three experimental settings—pure, dual-faction, and heterogeneous networks—we observed behavior-specific complexity patterns associated with nodes biased toward homophily, neophily, and social conformity. These patterns were often counterintuitive within the context of the model: homophilic nodes tended to exhibit increasing unpredictability over time; neophilic nodes displayed stable and bounded complexity despite continual novelty-seeking; and conformic nodes showed a two-phase pattern, with early predictability followed by later growth in unpredictability.

A key empirical result of this work is that the qualitative nLZ complexity patterns associated with different behavioral dispositions persist across diverse network compositions. More importantly, this consistency is complemented by explicit causal intervention tests. By altering the internal behavioral parameters of individual nodes while holding the surrounding social environment fixed, we demonstrate that changes in internal disposition induce systematic and directionally consistent shifts in the symbolic complexity of opinion trajectories. Within the context of this adaptive social network model, these intervention results provide direct evidence that internal behavioral rules exert a causal influence on the unpredictability of individual opinion trajectories, rather than merely correlating with favorable environmental conditions.

Supplementary permutation entropy analyses for Scenarios~1 and~3 confirm the presence of non-random ordinal structure but do not distinguish behavioral factions, reinforcing that the observed differences arise from long-range temporal organization captured by nLZ complexity (Figures S8-S11).

Within the limited but informative scope of this model, the results suggest an intriguing perspective on individual-level social experience. The complexity or volatility of an agent’s opinion trajectory appears to be closely linked to its own interaction rules, rather than being determined solely by the composition of the surrounding network. In this sense, the model raises the possibility that perceived instability or predictability in social experience may emerge endogenously from behavioral tendencies, rather than exclusively from external social conditions. While caution is essential when extrapolating from stylized models to real social systems, this perspective resonates with broader themes in psychology and social science concerning the interplay between individual dispositions and environmental influence. Rather than framing social instability purely as a consequence of external structure, the results highlight the potential importance of internal behavioral orientations in shaping subjective experience. The adaptive social network model studied here does not claim to capture human behavior in full complexity, but it offers a tractable setting in which such ideas can be systematically explored and refined.

Finally, the contribution of this work lies not only in the use of nLZ complexity, but in its emphasis on temporal, individual-level dynamical signatures within behaviorally heterogeneous adaptive networks. By combining observational analyses with targeted internal interventions, we demonstrate how behavioral dispositions can be causally linked to differences in opinion trajectory complexity within this modeling framework. Future work could extend this approach by incorporating complementary dynamical and information-theoretic measures—such as entropy-based metrics, recurrence quantification analysis, or fractal characteristics—to provide additional lenses on opinion dynamics, as each tool captures distinct aspects of temporal structure. At the same time, the explanations offered for the observed complexity trends remain phenomenological and conjectural; while causality is established at the level of internal parameters, a rigorous mechanistic derivation linking nLZ patterns to specific adaptive network processes—such as link weight evolution, clustering dynamics, or rewiring activity—remains an important direction for future work. More broadly, these results point toward a research direction in opinion dynamics that treats individual behavioral identity as a meaningful contributor to social complexity, while remaining attentive to the assumptions and limitations inherent in simplified models of social systems.

\bibliographystyle{elsarticle-num} 
\bibliography{References}

\end{document}